# Evaluation d'une requête en SQL

**Lect drd. Diana Sophia Codat**
"Tibiscus" University of Timisoara

**REZUMAT** Obiectivul acestei lucrari este de a arăta cum procesorul de interogări răspunde unei interogări SQL. Procesorul de interogări este descompus in două părți. Prima, numită compilare de interogare traduce o interogare SQL într-un plan de executare fizică A doua, numită evaluare de interogare execută planul de execuție.

## 1 Introduction

L'objectif de cet article est de montrer comment le processeur de requêtes répond à une requête SQL.

Le processeur de requête est décomposé en deux couches. La première appelée compilation d'une requête traduit une requête SQL en un plan d'exécution physique. La deuxième couche appelée évaluation d'une requête exécute ce plan d'exécution.

La compilation d'une requête consiste en trois étapes:
1. Analyse de la requête: le résultat est un plan d'exécution logique initial, représentation algébrique de la requête. L'interprétation de ce plan est une séquence d'opérations de l'algèbre relationnelle.
2. Réécriture de la requête: le plan initial est transformé en un plan équivalent grâce à des règles ou lois algébriques, plan qui est censé être plus efficace que le plan initial.
3. Génération d'un plan d'exécution physique: on sélectionne des algorithmes pour implanter chacune des opérations algébriques du plan logique et un ordre d'exécution. Le choix du meilleur plan dépend des opérations physiques implantant les divers opérateurs de l'algèbre, disponibles dans le processeur de requêtes. Il dépend également des *chemins d'accès* aux fichiers représentant les relations disponibles, c'est-à-dire de l'existence d'index ou de tables de hachage. Enfin il dépend aussi des données statistiques enregistrées ou estimées pour chaque relation (nombre de nuplets, de page d'une relation, sélectivité d'un attribut dans une relation, etc.). Le plan d'opérations physiques est également représenté sous forme d'arbre. Le plan comprend des détails tels que comment se fait l'accès à un fichier et si et quand une relation doit être triée.

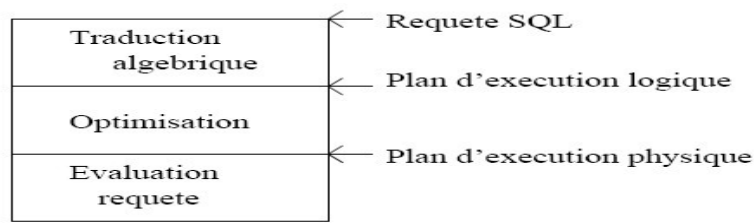

**Fig 1.** *Les étapes de traitement d'une requête*

Les informations telles que l'existence et la spécification d'index sur une table et les données statistiques enregistrées sont stockées dans le *catalogue* de la base. Celui-ci est également représenté sous forme de relations accédées et gérées par le système. Le catalogue est également appelé *schéma physique*. On dit qu'il contient des *méta-données*.

Les étapes 2 et 3 forment ce qu'on appelle l'*optimisation* d'une requête. Bien que nous distinguions ces deux étapes pour pédagogiquement mettre en évidence les différents types d'optimisation, il n'est pas clair que ces étapes soient séparées dans la sous couche optimisation d'un processeur de requête du commerce.

Ces différentes étapes seront étudiées dans la section 11.3. Avant de traiter la compilation d'une requête, on étudie en détail les différents algorithmes pour implanter les opérateurs de l'algèbre relationnelle.

## 2 Evaluation d'une requête

Le résultat de la compilation d'une requête est un plan d'exécution (physique), c'est-à-dire une séquence d'opérations à exécuter.

L'*évaluation d'une requête* consiste à exécuter cette séquence d'opérations:
- on accède aux fichiers, on fait une opération, on stocke éventuellement le résultat sur un fichier temporaire,

- on lance l'opération suivante. On présente éventuellement le résultat à l'utilisateur.

L'objectif de l'optimisation est une évaluation efficace de la requête initiale. Chaque opération est une implantation particulière d'une ou plusieurs opérations de l'algèbre relationnelle. Mais d'autres opérateurs sont nécessaires comme l'accès aux données, et en particulier le balayage d'une table, c'est-à-dire la lecture en mémoire centrale et le parcours d'une relation, nuplet par nuplet.

Le tri est une autre opération fondamentale pour l'évaluation des requêtes. On a besoin du tri par exemple lorsqu'on fait une projection ou une union et qu'on désire éliminer les nuplets dupliqués.

## 3 Techniques d'accès et prétraitement

Toute opération a une ou deux tables en entrée. La façon dont on accède aux nuplets de ces tables (*chemin d'accès*) a un impact sur l'efficacité de l'opération. Par ailleurs un prétraitement des tables avant l'exécution même de l'opération peut améliorer considérablement les performances de l'algorithme. On regroupe ces techniques d'accès et de prétraitement en quatre catégories.

1. *Accès séquentiel ou itération*. Tous les nuplets d'une table sont examinés lors de l'opération, en général suivant l'ordre dans lequel ils sont stockés.
2. *Accès par index*. Un sous-ensemble de nuplets sont accédés en deux étapes. La première consiste à traverser un index pour aller chercher les adresses de un ou plusieurs nuplets. La deuxième étape (*accès par adresse*) permet d'accéder aux nuplets connaissant leur adresse.
3. *tri*. Avant d'effectuer l'opération on trie la ou les tables. Cette technique est utilisée par exemple pour la projection et la jointure.
4. *hachage*. Avant d'effectuer l'opération on hache une table. Cette technique est par exemple utilisée pour la jointure.

Ces techniques seront introduites lors des algorithmes pour chaque opération. Quelques remarques sont nécessaires dès à présent:

Nous prenons comme hypothèse que les index s'ils existent sont denses et que les tables ne sont pas ordonnées suivant la clé d'index. Ce type d'organisation est souvent appelé dans la littérature anglophone "non clustered index".

– On pourrait rajouter une cinquième catégorie, accès à une table par son index de hachage. Nous supposons pour simplifier que les tables ne sont pas pré-hachées. Il n'existe pas d'index de hachage.

Cependant le hachage d'une table se fait en temps réel lors de l'exécution d'une opération (quatrième catégorie.)

Le tri et le hachage (catégories 3 et 4) sont deux techniques de prétraitement basées sur le *partitionnement* d'une table en fonction de la valeur d'une clé en vue de décomposer une opération en une collection d'opérations moins chères.

## 4. Amélioration de l'efficacité du parcours séquentiel

Comme le temps de réponse ou le temps d'évaluation est le critère d'efficacité et que celui-ci dépend du nombre de lectures disque, non seulement on réduit ce nombre, mais on essaie également de réduire le temps de lecture d'une page. Or très souvent, lorsqu'un nuplet est lu les nuplets suivants le sont ultérieurement.

Le balayage séquentiel d'une table en est un exemple. Les pages de la relation sont lues en séquence. Si ces pages sont contigues sur disque et et si on lit plusieurs pages contigues du disque à la fois, le temps de lecture est moindre que si on lit ces pages une à une (voir chapitre 1). D'où la nécessité
1. de regrouper les pages d'une table dans des espaces contigus appelés *extensions* ou *segments*,
2. et de faire une *lecture à l'avance*: quand on lit une page, on lit également les pages suivantes dans l'extension.

1. Pour simplifier on parle de table. En fait on devrait parler de la représentation physique d'une table, c'est-à-dire soit un fichier si la table est stockée sur disque, soit toute structure de données en mémoire centrale représentant un résultat intermédiaire.

Par conséquent l'unité d'entrée/sortie d'un SGBD (un bloc ou page) est souvent un multiple de celle du gestionnaire de fichier: la taille d'un tampon en mémoire est en général un multiple de la taille d'un bloc (ou page) physique.

*Le Tri externe*

Le tri d'une relation sur un ou plusieurs attributs utilise l'algorithme de tri-fusion. Celui-ci est du type "diviser pour régner" Il a deux phases:
1. Découpage de la table en partitions telles que chaque partition tienne en mémoire centrale et tri de chaque partition en mémoire. On utilise en général l'algorithme de *Quicksort*.
2. Fusion des partitions triées.

Regardons en détail chacune des phases.

***Phase de tri***

Supposons que nous disposons pour faire le tri de M tampons en mémoire. La relation est lue séquentiellement, M blocs par M blocs. Lorsque les M tampons ont été remplis par M blocs de la table, les nuplets de ces M tampons sont triés: ils forment alors une partition qui est écrite sur disque. A l'issue de cette phase on a B/M partitions triées, où B est le nombre de blocs de la relation.

### *Phase de fusion*

La phase de fusion consiste à récursivement fusionner les partitions. A chaque étape on obtient des partitions triées plus grosses jusqu'à ce qu' on obtienne la relation tout entière.

Commençons par regarder comment on fusionne en mémoire centrale deux listes triées A et B. On a besoin de trois tampons. Dans les deux premiers, les deux listes à trier sont stockées. Le troisième tampon sert pour le résultat c'est-à-dire la liste résultante triée. L'algorithme est donné ci-dessous:

> *a= premier élément de A;*
> *b= premier élément de B;*
> *tant qu'il reste un élément dans A ou B {*
> *si a avant b {*
> *si tampon sortie T plein {*
> *vider T*
> *}*
> *ecrire a dans T*
> *a= élément suivant dans A*
> *}*
> *sinon {*
> *si tampon sortie T plein {*
> *vider*
> *}*
> *ecrire b dans T*
> *b= élément suivant dans B*
> *}*
> *}*

***Algorithme fus: Fusion en mémoire de deux listes triées***

Par convention, lorsque tous les éléments de A(B) ont été lus, l'élément suivant est égal à *eof* qui est "après" tous les éléments de B(A).

## 5. Projection

S'il n'y a pas de sélection, la projection d'une relation consiste à examiner ses nuplets un par un, à effectuer sur chacun la projection, c'est-à-dire à éliminer les champs qu'on ne veut pas garder et enfin à éliminer les nuplets en double. Pour éliminer les dupliqués, il existe deux techniques, le tri et le hachage.

Dans le cas où les champs à projeter appartiennent à la clé d'un index de la table, il suffit de parcourir en séquence les feuilles de l'index. Celles-ci sont triées sur la clé. Il suffit de garder la clé ou son préfixe et d'éliminer les doublons voisins en séquence.

*Projection par tri*

L'algorithme comporte trois étapes:
1. Accès séquentiel de la table $R$ et projection de chaque nuplet. Le résultat est une table de taille $T$ pages.
2. Tri de cette table temporaire sur les attributs à projeter. Le résultat est une deuxième table intermédiaire de $T$ pages.
3. Parcours séquentiel de la table triée et en comparant les nuplets voisins, élimination des doublons.

On peut améliorer cet algorithme en faisant la projection et l'élimination des dupliqués en même temps que le tri:
1. Projeter les nuplets lors de la première phase du tri.
2. Eliminer les dupliqués lors des phases de fusions.

*Projection par hachage*

Soit M le nombre de tampons en mémoire centrale. L'algorithme est fait en deux phases: dans la première, la table est accédée séquentiellement et projetée en utilisant un seul des M tampons. Chaque nuplet projeté est placé dans l'un des M-1 tampons: la fonction de hachage h distribue uniformément les nuplets dans un entier compris entre 1 et M-1. Quand un tampon est plein il est stocké sur disque. A l'issue de cette phase on a M-1 partitions.

La deuxième phase consiste à éliminer les dupliqués. Elle repose sur le fait que si deux nuplets sont dupliqués ils sont dans la même partition.

**Bibliographie**